\pdfoutput=1 
\documentclass[showpacs,pra,aps,superscriptaddress,twocolumn,floatfix]{revtex4}
\usepackage{graphicx,float,amsmath,amssymb}
\usepackage{color}

\usepackage{mathrsfs} 
\usepackage[utf8]{inputenc}  
\usepackage[francais,english]{babel}

\usepackage{multirow} 

\usepackage{url,hyperref}  

\usepackage{wasysym}

\newfont{\ensmathquatorze}{msbm10 scaled 1400}
\newfont{\ensmathonze}{msbm10 scaled 1100}
\newfont{\ensmathdix}{msbm10}
\newfont{\ensmathneuf}{msbm10 scaled 833}
\newfont{\ensmathhuit}{msbm10 scaled 694}
\newfam\ensmathfam                        
\textfont\ensmathfam=\ensmathonze        
\scriptfont\ensmathfam=\ensmathdix       
\scriptscriptfont\ensmathfam=\ensmathhuit

\renewcommand{\leq}{\leqslant}
\renewcommand{\geq}{\geqslant}

\newcommand{\ket}[1]{|\kern.3ex#1\kern.3ex\rangle}
\newcommand{\bra}[1]{\langle\kern.3ex #1 \kern.3ex|}
\newcommand{\scalar}[2]{\langle\kern.3ex #1 \kern.3ex|\kern.3ex#2\kern.3ex\rangle}
\newcommand{\cum}[1]{\langle\langle #1 \rangle\rangle} 

 \newcommand{\EXP}[1]{{\mathrm{e}}^{#1}}         

\newcommand{\heaviside}{\mathop{\theta_\mathrm{H}}\nolimits}  

\def\D{\mathrm{d}}                  

\newcommand{\deriv}[2]{\frac{\mathrm{d}#1}{\mathrm{d}#2}}

\def\es{e} 
\def\hs{h} 
\def\ps{p} 

\def\kB{k_B}

\def\can{{\mathrm{c}}}
\def\gc{{\mathrm{g}}}

\def\XiB{\Xi^\mathrm{B}}
\def\XiF{\Xi^\mathrm{F}}

\def\TB{T_\mathrm{B}}
\def\cum{c}

\def\Gf{G}

\begin{document}

\selectlanguage{english}

\title{Correlations of occupation numbers in the canonical ensemble \\ and application to BEC in a 1D harmonic trap}

\author{Olivier Giraud}
\affiliation{LPTMS, CNRS, Univ.~Paris-Sud, Universit\'e Paris-Saclay, F-91405 Orsay, France}

\author{Aur\'elien Grabsch}
\affiliation{LPTMS, CNRS, Univ.~Paris-Sud, Universit\'e Paris-Saclay, F-91405 Orsay, France}

\author{Christophe Texier}
\affiliation{LPTMS, CNRS, Univ.~Paris-Sud, Universit\'e Paris-Saclay, F-91405 Orsay, France}

\date{November 28, 2018}

\begin{abstract}
We study statistical properties of $N$ non-interacting identical bosons or fermions in the canonical ensemble.
We derive several general representations for the $p$-point correlation function of occupation numbers $\overline{n_1\cdots n_p}$.
We demonstrate that it can be expressed as a ratio of two $p\times p$ determinants involving the (canonical) mean occupations $\overline{n_1}$, ..., $\overline{n_p}$, which can themselves be conveniently expressed in terms of the $k$-body partition functions (with $k\leq N$).
We draw some connection with the theory of symmetric functions, and obtain an expression of the correlation function in terms of Schur functions. 
Our findings are illustrated by revisiting the problem of Bose-Einstein condensation in a 1D harmonic trap, for which we get analytical results. 
We get the moments of the occupation numbers and the correlation between ground state and excited state occupancies.
In the temperature regime dominated by quantum correlations, the distribution of the ground state occupancy is shown to be a truncated Gumbel law.
The Gumbel law, describing extreme value statistics, is obtained when the temperature is much smaller than the Bose-Einstein temperature.
\end{abstract}



\maketitle


\section{Introduction}

The theory of non-interacting identical quantum particles is a fundamental block of the basic education in statistical physics \cite{Hua63,PatBea11,TexRou17book}. 
In the standard approach, calculations are performed in the grand canonical ensemble, as it provides the clearest and most efficient tools to relate single-particle and thermodynamic properties. One can then use the equivalence between statistical physics ensembles in the thermodynamic limit in order to get the thermodynamic observables as a function of the relevant parameters (energy or temperature, number of
particles or chemical potential, etc). One should however keep in mind that the correspondence between ensembles in the thermodynamic limit only holds for \textit{averages} of observables, and not for their fluctuations \cite{PatBea11,TexRou17book}.

Recently, remarkable progress in atomic physics of ultracold atoms has raised many questions concerning many-body effects in those systems \cite{BloDalZwe08}. 
Simpler questions related to quantum correlations in non-interacting gases have also been put forward, as experiments deal with extremely diluted gases, for which the non-interacting limit is often a good starting point. 
Depending whether we deal with bosons or fermions, the situation can be rather different. 
This difference manifests itself, for example, in the presence of a harmonic confinement, as is realised in experiments with optical traps. 
For bosonic gases, although the low temperature properties are dominated by interactions, many basic properties like energy or density profile can then be obtained within a mean field approximation~\cite{DalGioPitStr99}. 
In fermionic gases, the Pauli principle strongly suppresses the effect of interactions at low temperature, which makes the non-interacting description a good starting point, from which interaction can be treated perturbatively~\cite{GioPitStr08}.
Due to cooling techniques by evaporation, a trapped ultracold atomic gas only contains a moderately large number of atoms (few thousands to few millions), 
what has led, before considering interaction effects, to re-examine the differences between the various statistical physics ensembles for non-interacting particles~\cite{BroDevLem96,SchMed96,GroHol96,WeiWil97,WilWei97,GajRza97,NavBitGajIdzRza97,HolKalKir98,ChaMekZam99,Pra00}, as the microcanonical or canonical ensembles are more relevant in this case (see the review~\cite{DalGioPitStr99}).

\subsection{Occupation numbers}

Not only thermodynamic properties and global observables are of interest, but, motivated by the remarkable achievement of the ``atomic microscope'' \cite{CheNicOkaGerRamBakWasLomZwi15,HalHudKelCotPeaBruKuh15,ParHubMazChiSetWooBlaGre15}, also local observables have been studied recently (for a review see Ref.~\cite{DeaLeDMajSch16}).
A basic ingredient of such studies is the knowledge of the number of particles $n_\lambda$ in each individual eigenstate $\ket{\lambda}$.
The \textit{grand canonical} mean occupation is given by the Bose-Einstein or Fermi-Dirac distribution
\begin{equation}
  \label{eq:BEandFD}
  \overline{n_\lambda}^{\gc}  =  \frac{1}{1/(x_\lambda\varphi)\mp1},
  \qquad  
  x_\lambda = \EXP{-\beta\varepsilon_\lambda}
  \:,
\end{equation}
where $\varphi$ is the fugacity, $\{\varepsilon_\lambda\}$ are the individual energy levels and $\beta=1/(\kB T)$ is the inverse temperature. $\overline{\cdots}^{\gc}$ is the grand canonical average.
In this formula, the upper ($-$) and lower ($+$) signs stand for bosons and fermions, respectively. 
For a fixed number $N$ of particles, the \textit{canonical} mean occupation numbers can be expressed in terms of the $k$-body canonical partition functions with $k=1,\ldots,N$. The canonical partition function $Z_N(\beta)$ for $N$ bosons or fermions can be obtained by recursion from the formula~\cite{Lan61,For71,BorFra93,SchSch02,MulFer03}
\begin{equation}
  \label{eq:Landsberg1961}
  Z_N(\beta)
  = \frac{1}{N} \sum_{k=1}^N 
  (\pm1)^{k-1} \, Z_1(k\beta)\,Z_{N-k}(\beta)
  \:,
\end{equation}
where $Z_1(\beta)=\sum_\lambda \EXP{-\beta\varepsilon_\lambda}$ is the canonical partition function for one particle, and the upper ($+$) and lower ($-$) signs stand for bosons and fermions, respectively. The canonical mean occupation number is then given by 
\begin{equation}
  \label{eq:BEandFDcanonical}
  \overline{n_\lambda} 
  = \sum_{k=1}^N (\pm1)^{k-1} \frac{Z_{N-k}}{Z_N} \, x_\lambda^{k}
\end{equation}
with $Z_0=1$ \cite{WeiWil97,BorHarMulHil99,ChaMekZam99}. 
The canonical average is simply denoted $\overline{\cdots}$.
This expression is expected to coincide with \eqref{eq:BEandFD} in the thermodynamic limit, provided that the fugacity $\varphi$ in Eq.~\eqref{eq:BEandFD} is chosen in such a way that the condition $\sum_\lambda\overline{n_\lambda}^{\gc}=N$ is fulfilled (for a quantitative discussion, cf.~Ref.~\cite{GraMajSchTex18} where it was shown that $\overline{n_\lambda} - \overline{n_\lambda}^{\gc}=\mathcal{O}(N^{-1})$ for trapped fermions in one dimension).
The power of the grand canonical ensemble lies in the \textit{independence} of individual energy level properties (which leads to many useful additivity properties for the thermodynamic observables), i.e. the absence of correlations between occupation numbers
\begin{equation}
  \label{eq:TrivialUncorrelationsGC}
    \overline{n_{\lambda_1}\cdots n_{\lambda_p}}^\gc
  = \overline{n_{\lambda_1}}^\gc\times\cdots\times\overline{n_{\lambda_p}}^\gc
  \:.
\end{equation}
However, in the canonical ensemble, the constraint on the total number of particles $\sum_\lambda n_\lambda = N$ implies non-trivial correlations between occupation numbers, $\overline{n_1n_2}\neq\overline{n_1}\times\overline{n_2}$. In the present paper, we focus on the canonical ensemble and study these correlations.

\subsection{Main results}

Our main results are complementary expressions for the correlation function. In order to lighten notations, we consider the correlations of the first $p$ levels~; this does not imply any restriction on the generality of the discussion, as levels all play an equivalent role. The first expression we obtained is in terms of the canonical partition functions and of the $x_\lambda = \EXP{-\beta\varepsilon_\lambda}$, and reads
\begin{align}
  \label{eq:TheMainResult0}
  \overline{n_1\cdots n_p} 
  &= \frac{(\pm1)^p}{Z_N}
  \\
  \nonumber
  &\times
  \hspace{-1cm}
  \sum_{
     {\scriptsize
       \begin{array}{c}
          k_1,\ldots,\, k_p=1 \\ k_1+\cdots+k_p\leq N 
       \end{array}
       }
       } ^N
  \hspace{-1cm}
  (\pm x_1)^{k_1}\cdots(\pm x_p)^{k_p}\, Z_{N-k_1-\cdots-k_p}
   \:.
\end{align}
This generalizes \eqref{eq:BEandFDcanonical}.
The second expression is a representation in terms of two $p\times p$ determinants,
\begin{equation}
  \label{eq:TheMainResult}
    \overline{n_1n_2\cdots n_p} 
    = 
  (\mp1)^{p-1}
    \frac{ 
    \left|\begin{array}{cccc}
     \overline{n_1} & x_1 & \cdots & x_1^{p-1} \\
     \overline{n_2} & x_2 & \cdots & x_2^{p-1} \\
     \vdots & \vdots & \ddots & \vdots \\
     \overline{n_p} & x_p & \cdots & x_p^{p-1}
    \end{array}\right|
    }{
    \left|\begin{array}{cccc}
     1 & x_1 & \cdots & x_1^{p-1} \\
     1 & x_2 & \cdots & x_2^{p-1} \\
     \vdots & \vdots & \ddots & \vdots \\
     1 & x_p & \cdots & x_p^{p-1}
    \end{array}\right|
    }
    \:,
\end{equation}
where  the denominator is a Vandermonde determinant. A remarkable observation is that the $p$-point correlation function can be expressed only in terms of the mean values $\overline{n_1},\:\ldots,\overline{n_p}$. 
It can also be expressed in terms of $q$-point functions with $q<p$, see Eq.~\eqref{eq:geneqarb} below.
Equation~\eqref{eq:TheMainResult} turns out to be valid also in the grand canonical ensemble. 
A particular instance of this general relation for $p=2$,
\begin{equation}
  \label{eq:UsefulRecurrence}
        \overline{n_1n_2} = \mp\, 
      \frac{ \EXP{\beta\varepsilon_1}\,\overline{n_1} - \EXP{\beta\varepsilon_2}\,\overline{n_2} }
           { \EXP{\beta\varepsilon_1} - \EXP{\beta\varepsilon_2} }
           \:,
\end{equation}
was recently used for the study of fluctuations of certain observables for trapped non-interacting one-dimensional fermions \cite{GraMajSchTex18}.
Equation~\eqref{eq:TheMainResult0} and Eq.~\eqref{eq:UsefulRecurrence} were obtained very recently in Ref.~\cite{Sch17} in the fermionic case.

In the case of bosons, we could express the general correlation function for arbitrary integer powers $r_1,\:\ldots,r_p$ as
\begin{align}
  \label{eq:GeneralCorrelations}
  &\overline{n_1^{r_1}\cdots n_p^{r_p}}  
  = \frac{1}{Z_N}
  \\
  \nonumber
  &\times
  \hspace{-1cm}
  \sum_{
     {\scriptsize
       \begin{array}{c}
          k_1,\,\ldots,\, k_p=1 \\ \ k_1+\cdots+k_p\leq N 
       \end{array}
       }
       } ^N
  \hspace{-1cm}
  a_{k_1}(r_1)\,x_1^{k_1}\cdots a_{k_p}(r_p)\,x_p^{k_p}\, Z_{N-k_1-\cdots-k_p}
  \:,
\end{align}
where
\begin{equation}
  a_k(r)=k^r-(k-1)^r
  \:.
\end{equation}
Equation \eqref{eq:GeneralCorrelations} generalizes~\eqref{eq:TheMainResult0}.

We illustrate these formulae by considering the study of Bose-Einstein condensation in a one-dimensional harmonic trap. We obtain several simple analytical results for the occupation numbers of single-particle levels. 
In particular, we have obtained the distribution $\mathscr{P}_{k,N}$ of the occupation number $n_k$, with $k\in\mathbb{N}$, for $N$ bosons in the canonical ensemble. 
In the quantum regime $\omega \ll T\ll N\omega$, where $\omega$ is the trap frequency, we have obtained the scaling form
\begin{equation}
  \label{eq:IntroPk}
  \mathscr{P}_{k,N}(n) \simeq \frac{\omega}{T}\, Q_{k,z}\!\left( \frac{\omega\,n}{T} \right)
\end{equation}
where 
\begin{equation}
  \label{eq:IntroQk}
  Q_{k,z}(\xi) = \heaviside(\xi)\,\EXP{z}
  \left(k+z\EXP{\xi}\right)\,\exp\left\{-k\xi-z\EXP{\xi}\right\}
  \:,
\end{equation}
with $z=(T/\omega)\,\exp[-N\omega/T]$ and $\heaviside$ the Heaviside step function.
This expression can be compared with the similar distribution obtained in the grand canonical ensemble
$\mathscr{P}_{k}^\gc(n)\propto\varphi^n\EXP{-n\beta\varepsilon_k}$, which would give, after the same rescaling as above, $Q_{k}^\gc(\xi)\propto\varphi^{T\xi/\omega}\,\EXP{-k\xi}$.
The case of the ground state is of special interest:
 \eqref{eq:IntroPk} and \eqref{eq:IntroQk} simplify as
\begin{align}
  &\mathrm{Proba}\left\{ n_0 = N - \frac{T\ln(T/\omega)}{\omega} + \frac{T}{\omega}\zeta \right\}
  \nonumber\\
  &\hspace{2cm}\propto 
  \heaviside(\zeta-\ln z)\, \exp\left\{\zeta - \EXP{\zeta} \right\}
  \:,
\end{align}
which corresponds to a truncated Gumbel law.

\subsection{Outline}

The outline of the article is as follows~:
in Section~\ref{sec:SymmetricPolynomials}, we recall the connection between our problem and the theory of symmetric functions, which will allow us to introduce some useful tools. 
Our main results expressing correlations between occupation numbers, Eqs.~\eqref{eq:TheMainResult0} and \eqref{eq:TheMainResult}, are derived in Section~\ref{sec:Correlations}. Finally, in Section~\ref{sec:BEC1D}, we illustrate our results on the problem of Bose-Einstein condensation in a one-dimensional harmonic trap.


\section{Symmetric functions}
\label{sec:SymmetricPolynomials}

The connection between the problem of identical particles and the theory of symmetric functions has been discussed in Refs.~\cite{For71,Bal01}. 
In Ref.~\cite{SchSch02}, Schmidt and Schnack have pointed out that the relation \eqref{eq:Landsberg1961} for fermions, which is attributed to Landsberg~\cite{Lan61} in many articles, is nothing else but the well-known Newton identity. 
In this section, we introduce some useful notation. 
As a simple illustration, we will recover the relations \eqref{eq:Landsberg1961}. 
For a recent reference on the mathematical theory of symmetric functions, see the monograph~\cite{Mac95}.

\subsection{Families of symmetric polynomials}

A function $\phi$ in $M$ variables $x_1,\,x_2,\ldots,x_M$ is said to be symmetric if $\phi(x_{\sigma(1)},\ldots, x_{\sigma(M)}) = \phi(x_1,\ldots, x_M)$ for any permutation $\sigma$ of the $M$ indices. We now introduce three useful families of symmetric polynomials.
\\

The {\it elementary symmetric polynomials} are defined as
\begin{equation}
    \es_{N}(x_1,\ldots,x_M) = \sum_{\lambda_1<\lambda_2<\cdots<\lambda_N}\hspace{-.5cm} x_{\lambda_1}x_{\lambda_2}\cdots x_{\lambda_N},
\end{equation}
with $\es_0 = 1$ and $\es_k = 0$ for  $k>M$. For example, $\es_1(x_1,\ldots,x_M) = x_1 + x_2 + \cdots + x_M$ and $\es_2(x_1,\ldots,x_M) =  x_1x_2 + x_1x_3 + \cdots$. Their generating function is given by
\begin{equation}
  \label{eq:GPFfermions}
  \XiF(\varphi) 
  = \sum_{N=0}^M \es_N\,\varphi^N 
  = \prod_{\lambda=1}^M (1+\varphi\,x_\lambda)
  \:.
\end{equation}

The {\it complete homogeneous symmetric polynomials} are defined as
\begin{equation}
    \hs_{N}(x_1,\ldots,x_M) = \sum_{\lambda_1\leq \lambda_2\leq \cdots\leq \lambda_N}\hspace{-.5cm}  x_{\lambda_1}x_{\lambda_2}\cdots x_{\lambda_N},
\end{equation}
with $h_0=1$. For example, 
$
  \hs_3(x_1,x_2,x_3) 
  = (x_1^3+x_2^3+x_3^3)
  + (x_1^2x_2+x_1^2x_3+\cdots)
  + (x_1x_2x_3)
$.
Their generating function is given by
\begin{equation}
  \label{eq:GPFbosons}
  \XiB(\varphi) 
  = \sum_{N=0}^\infty \hs_N\,\varphi^N 
  = \prod_{\lambda=1}^M (1-\varphi\,x_\lambda)^{-1}
  \:.
\end{equation}
Note that, contrary to the sum in the definition of $\XiF(\varphi)$, the sum here extends to infinity, as $\hs_N\neq0$ for $N>M$.

The {\it power sum polynomials}  are defined as
\begin{equation}
  \ps_k(x_1,\ldots,x_M) = \sum_{\lambda=1}^M x_\lambda^k
  \hspace{0.5cm}\mbox{for }
  k\geq1
  \:.
\end{equation}
Their generating function is given by
\begin{equation}
  P(\varphi) 
  = \sum_{k=1}^\infty \ps_k\,\varphi^{k-1} 
  = \sum_{\lambda=1}^M \frac{x_\lambda}{1-\varphi\,x_\lambda}
\end{equation}
(for convenience the definition is slightly different from those of the previous generating functions as we did not introduce a $\ps_0$).

\subparagraph{Correspondence with the problem of identical particles.---}

Let us consider $N$ particles in $M$ energy levels (possibly infinite). Setting 
$x_\lambda=\EXP{-\beta\varepsilon_\lambda}$ as in Eq.~\eqref{eq:BEandFD}, one readily sees that the canonical partition function for bosons $Z^\mathrm{B}_N(\beta)$ coincides with $h_N$, while 
the canonical partition function for fermions $Z^\mathrm{F}_N(\beta)$ coincides with $e_N$. Moreover, one obviously has $\ps_1=\hs_1=\es_1$, and $\ps_k(x_1,\ldots,x_M) = \es_1(x_1^k,\ldots,x_M^k) = \hs_1(x_1^k,\ldots,x_M^k)$, so that the single-particle canonical partition function at inverse temperature $k\beta$, which is $Z^\mathrm{B}_1(k\beta)$ or $Z^\mathrm{F}_1(k\beta)$, coincides with $p_k$. 
One can thus establish the following dictionary between the mathematician's and the physicist's notations~:
\begin{align*}
   x_\lambda & \longrightarrow \EXP{-\beta\varepsilon_\lambda}
   \\
   \es_N & \longrightarrow Z^\mathrm{F}_N(\beta)
   \\
   \hs_N & \longrightarrow Z^\mathrm{B}_N(\beta)
   \\
   \ps_k & \longrightarrow Z^\mathrm{B}_1(k\beta)=Z^\mathrm{F}_1(k\beta).
\end{align*}
The generating functions $\XiB(\varphi)$ and $\XiF(\varphi)$ coincide with the grand canonical partition functions for bosons and fermions, respectively.

\subsection{Newton identity}

There exist various identities relating the generating functions $\XiB(\varphi)$, $\XiF(\varphi)$ and $P(\varphi)$. Expanding these identities in powers of $\varphi$ provides some relations between the three families of symmetric polynomials defined above. For instance, the duality relation 
\begin{equation}
  \label{eq:DualityBosonFermion}
  \XiB(\varphi)\,\XiF(-\varphi) = 1
  \:,
\end{equation}
readily obtained from \eqref{eq:GPFfermions} and \eqref{eq:GPFbosons}, allows to express the $\es_k$'s in terms of the $\hs_k$'s, or conversely.

Another identity that can be easily obtained from the expressions of the previous subsection is 
\begin{equation}
  \label{eq:RelationGF}
   P(\varphi) =  \deriv{}{\varphi} \ln \XiB(\varphi) = - \deriv{}{\varphi} \ln \XiF(-\varphi) 
   \:,
\end{equation}
that is,
\begin{equation}
  \label{eq:RelationGFf}
  -\deriv{}{\varphi} \XiF(-\varphi)=P(\varphi)\,\XiF(-\varphi)
\end{equation}
and 
\begin{equation}
 \label{eq:RelationGFb}
  \deriv{}{\varphi} \XiB(\varphi)=P(\varphi)\,\XiB(\varphi)
  \:.
\end{equation}
Expanding explicitly \eqref{eq:RelationGFf} in powers of $\varphi$ yields
\begin{equation}
 \sum_{N=1}^M  N\,(-\varphi)^{N-1}  \, \es_N
 = 
 \sum_{k=1}^\infty \ps_k\, \varphi^{k-1} \sum_{j=0}^M (-\varphi)^j\, \es_j
 \:.
\end{equation}
Identification of terms in $\varphi^{N-1}$ in the r.h.s. gives the relation
\begin{equation}
  \label{eq:GirardNewton}
    \es_N = \frac{1}{N} \sum_{k=1}^N (-1)^{k-1} \, \ps_k\, \es_{N-k}
    \:.
\end{equation}
This is precisely Eq.~\eqref{eq:Landsberg1961} for fermions, according to the above dictionary. 
The relation \eqref{eq:GirardNewton} was derived by Isaac Newton in his book, published in 1666 (see Ref.~\cite{Whi67}, p.~519). It is known as the Newton identity \cite{Mac95}.
It is interesting to point that similar identities, expressing the $\ps_k$'s in terms of the $\es_k$'s, 
were obtained earlier by Albert Girard in 1629 (see Ref.~\cite{Gir1629} page F2, i.e. page~$\sim50$ of the manuscript, where the elementary polynomial $\es_k$ is called ``$k$-th mesl\'e'').
Girard's identities can be obtained by identifying the terms $\varphi^{N-1}$ of 
$P(-\varphi)=\deriv{}{\varphi}\ln\Xi^\mathrm{F}(\varphi)$ 
(the cases $N=1,\,2,\,3$ and $4$ are considered in this old manuscript).
 
Expanding \eqref{eq:RelationGFb} in powers of $\varphi$ gives
\begin{equation}
 \sum_{N=1}^\infty N\,\varphi^{N-1} \, \hs_N
 = 
 \sum_{k=1}^\infty \ps_k\, \varphi^{k-1} \sum_{j=0}^\infty \varphi^j\, \hs_j,
\end{equation}
leading to the relation
\begin{equation}
  \label{eq:BosonicNewton}
    \hs_N = \frac{1}{N} \sum_{k=1}^N  \ps_k\, \hs_{N-k}
    \:.
\end{equation}
This corresponds to Eq.~\eqref{eq:Landsberg1961} for bosons.

The theory of symmetric functions also provides a determinantal representation of elementary symmetric polynomials in terms of power sum polynomials 
(p.~28 of \cite{Mac95}), as 
\begin{equation}
  \label{eq:ZNdetFerm}
  \es_N 
  = \frac{1}{N!}
  \left|
    \begin{array}{ccccc}
      \ps_1     & 1         & 0         & \cdots  & 0       \\
      \ps_2     & \ps_1     & 2         &         & 0       \\
      \vdots    & \vdots    & \ddots    & \ddots  & \vdots  \\
      \ps_{N-1} & \ps_{N-2} &           &         & N-1     \\
      \ps_{N}   & \ps_{N-1} & \ps_{N-2} & \cdots  & \ps_1 
    \end{array}
  \right|
  \:.
\end{equation}
This provides an explicit expression of the $N$-body partition function $Z^\mathrm{F}_N(\beta)$ in terms of the one particle partition function $Z_1(k\beta)$. The homogeneous polynomials $\hs_N$ can also be expressed by the r.h.s.~of Eq.~\eqref{eq:ZNdetFerm}, but with the determinant replaced by a permanent \cite{For71}. 
Alternatively, they can be expressed in terms of a determinant \cite{Mac95}, as
\begin{equation}
  \label{eq:ZNdetBos}
  \hs_N 
  = \frac{1}{N!}
  \left|
    \begin{array}{ccccc}
      \ps_1     & -1         & 0         & \cdots  & 0       \\
      \ps_2     & \ps_1     & -2         &         & 0       \\
      \vdots    & \vdots    & \ddots    & \ddots  & \vdots  \\
      \ps_{N-1} & \ps_{N-2} &           &         & -N+1     \\
      \ps_{N}   & \ps_{N-1} & \ps_{N-2} & \cdots  & \ps_1 
    \end{array}
  \right|
  \:,
\end{equation}
 which provides an expression of $Z^\mathrm{B}_N(\beta)$ in terms of~$Z_1(k\beta)$.


\section{Correlation functions}
\label{sec:Correlations}

The tools developed in the previous section allow us to easily derive Eqs.~\eqref{eq:TheMainResult0} and \eqref{eq:TheMainResult}, as we now show.

\subsection{Canonical and grand canonical ensembles}

We consider $N$ identical particles in $M$ (possibly infinite) energy levels.
The occupation number of the individual eigenstate $\ket{\lambda}$ of energy $\varepsilon_\lambda$ is denoted by $n_\lambda$.
Thus we have $n_\lambda\in\{0,\,1\}$ for fermions and $n_\lambda\in\mathbb{N}$ for bosons.
A basis of Fock space is given by the many-body states $\ket{\{n_\lambda\}}$, which are fully specified by the knowledge of all occupation numbers. We can express the canonical (Gibbs) distribution at inverse temperature $\beta$ as
\begin{equation}
  \label{eq:DistribCano}
    \mathscr{P}_N^\can(\{n_\lambda\})
    = \frac{\prod_\lambda x_\lambda^{n_\lambda}}{Z_N}\, 
    \delta_{N,\sum_\lambda n_\lambda},
\qquad
  x_\lambda=\EXP{-\beta\varepsilon_\lambda}
  \:,
\end{equation}
which gives the probability of occupying the quantum state $\ket{\{n_\lambda\}}$. Here $Z_N$ is the $N$-body canonical partition function, and the Kronecker symbol constrains the number of particles to be $N$. On the other hand, the grand canonical distribution is controlled by the fugacity $\varphi$, and reads
\begin{equation}
  \label{eq:DistribGrandCano}
    \mathscr{P}^\gc(\{n_\lambda\};\varphi)
    = \frac{\prod_\lambda (x_\lambda\varphi)^{n_\lambda}}{\Xi(\varphi)}\, 
\end{equation}
with $\Xi(\varphi)=\prod_\lambda(1\mp\varphi x_\lambda)^{\mp1}$ the grand canonical partition function 
given by Eqs.~\eqref{eq:GPFfermions} or \eqref{eq:GPFbosons}. 
For bosons, the convergence of the series is ensured by the condition $\varphi x_0<1$, where $\varepsilon_0$ is the individual ground state ($x_0=\EXP{-\beta\varepsilon_0}$).

Using these distributions, one can relate the grand canonical average $\overline{\cdots}^\gc$ and the canonical average $\overline{\cdots}^{(N)}$ for $N$ particles (the superscript will only be introduced if needed). Indeed, if $\mathcal{A}(\cdot)$ is any function of the occupation numbers, then, from \eqref{eq:DistribCano}--\eqref{eq:DistribGrandCano} one has
\begin{equation}
  \label{eq:GCandCANmeans}
  \Xi(\varphi)\,  \overline{\mathcal{A}(\{n_\lambda\})}^\gc 
  = \sum_{N=0}^\infty Z_N\, \overline{\mathcal{A}(\{n_\lambda\})}^{(N)}\, \varphi^N.
\end{equation}

\subsection{$p$-point correlation functions}

\subsubsection{Proof of Eqs.~\eqref{eq:TheMainResult0} and \eqref{eq:TheMainResult}}

We now apply \eqref{eq:GCandCANmeans} to $\mathcal{A}(\{n_\lambda\})=n_1\cdots n_p$. In the grand canonical ensemble, the occupation numbers are independent (see Eq.~\eqref{eq:TrivialUncorrelationsGC}), and they are given by Eq.~\eqref{eq:BEandFD}. We thus get from \eqref{eq:GCandCANmeans}
\begin{align}
  &\sum_{N=p}^\infty
  Z_N\,\overline{n_1\cdots n_p}^{(N)}
  \,\varphi^N
  \nonumber\\
  \label{eq:BeautifulGeneratingFunction}
  = &\frac{\Xi(\varphi)}{[1/(x_1\,\varphi)\mp1]\cdots[1/(x_p\,\varphi)\mp1]}
  \:.
\end{align}
The expansion of \eqref{eq:BeautifulGeneratingFunction} and the identification of each power of $\varphi$ directly gives Eq.~\eqref{eq:TheMainResult0}.
Consider for example $p=1$. We have explicitly
\begin{equation}
  \sum_{N=1}^\infty Z_N\,\overline{n_1}^{(N)}\,\varphi^N
  = x_1\,\varphi\sum_{k=0}^\infty (\pm x_1\,\varphi)^k \sum_{m=0}^\infty Z_m\,\varphi^m
  \:.
 \label{eq:BeautifulGeneratingFunctionp1}
\end{equation}
Identification of the $\varphi^N$ terms on both sides gives $Z_N\,\overline{n_1}^{(N)} =\sum_{q=1}^N (\pm1)^{q-1} x_1^q \, Z_{N-q}$, which is Eq.~\eqref{eq:BEandFDcanonical}.

\begin{widetext} 
We now introduce the $p\times p$ determinant
\begin{align}
  \label{eq:StartingPointDem}
  \sum_{N=1}^\infty
  Z_N\,
  \left|
  \begin{array}{ccccc}
      \overline{n_1}^{(N)} & x_1 & x_1^2 & \cdots & x_1^{p-1}\\
      \overline{n_2}^{(N)} & x_2 & x_2^2 & \cdots & x_2^{p-1}\\
      \vdots & \vdots & \vdots & \ddots & \vdots\\
      \overline{n_p}^{(N)} & x_p & x_p^2 & \cdots & x_p^{p-1}      
  \end{array}
  \right|
  \,\varphi^N
  =
  \left|
  \begin{array}{ccccc}
      \sum_NZ_N\,\overline{n_1}^{(N)}\,\varphi^N \, & x_1 & x_1^2 & \cdots & x_1^{p-1} \\
      \sum_NZ_N\,\overline{n_2}^{(N)}\,\varphi^N \, & x_2 & x_2^2 & \cdots & x_2^{p-1} \\
      \vdots & \vdots & \vdots &  \ddots& \vdots \\
      \sum_NZ_N\,\overline{n_p}^{(N)}\,\varphi^N \, & x_p & x_p^2 & \cdots & x_p^{p-1}        
  \end{array}
  \right|.
\end{align}
Inserting \eqref{eq:BeautifulGeneratingFunctionp1} in the right-hand side of this expression, we get
\begin{equation}
 \label{eq:MiddlePointDem}
 \left|
  \begin{array}{ccccc}
      \frac{x_1\,\varphi\,\Xi(\varphi)}{1\mp x_1\varphi} & x_1  & x_1^2 & \cdots & x_1^{p-1} \\
      \frac{x_2\,\varphi\,\Xi(\varphi)}{1\mp x_2\varphi} & x_2  & x_2^2 & \cdots & x_2^{p-1} \\
      \vdots                                             &\vdots&\vdots & \ddots & \vdots    \\
      \frac{x_p\,\varphi\,\Xi(\varphi)}{1\mp x_p\varphi} & x_p  & x_p^2 & \cdots & x_p^{p-1}        
  \end{array}
  \right|
=
 \frac{x_1\cdots x_p\,\varphi\,\Xi(\varphi)}{(1\mp x_1\varphi)\cdots(1\mp x_p\varphi)}
  \left|
  \begin{array}{ccccc}
      1 & (1\mp x_1\varphi) & x_1(1\mp x_1\varphi) & \cdots & x_1^{p-1}(1\mp x_1\varphi) \\
      1 & (1\mp x_2\varphi) & x_2(1\mp x_2\varphi) & \cdots & x_2^{p-1}(1\mp x_2\varphi) \\
      \vdots & \vdots & \vdots  &  \ddots & \vdots \\
      1 & (1\mp x_p\varphi) & x_p(1\mp x_p\varphi) & \cdots & x_p^{p-1}(1\mp x_p\varphi)        
  \end{array}
  \right|.
  \end{equation}
From linear combinations of columns in the determinant we then obtain
\begin{equation}
  \label{eq:EndPointDem}
 \frac{x_1\cdots x_p\,\varphi\,\Xi(\varphi)}{(1\mp x_1\varphi)\cdots(1\mp x_p\varphi)}
  (\mp\varphi)^{p-1}
  \left|
  \begin{array}{ccccc}
      1 &  x_1  & x_1^2 & \cdots & x_1^{p-1} \\
      1 & x_2 & x_2^2 & \cdots & x_2^{p-1} \\
      \vdots & \vdots &\vdots & \ddots & \vdots\\
      1 & x_p & x_p^2 & \cdots   & x_p^{p-1}      
  \end{array}
  \right|
  =
  (\mp1)^{p-1}
  \left|
  \begin{array}{ccccc}
      1 &  x_1  & x_1^2 & \cdots & x_1^{p-1}\\
      1 & x_2 & x_2^2 & \cdots & x_2^{p-1}\\
      \vdots & \vdots & \vdots & \ddots &\vdots\\
      1 & x_p & x_p^2 & \cdots  & x_p^{p-1}      
  \end{array}
  \right|
  \sum_{N=p}^\infty
  Z_N\,\overline{n_1\cdots n_p}^{(N)}
  \,\varphi^N,
\end{equation}
where we have used \eqref{eq:BeautifulGeneratingFunction} for the last equality. Identifying terms in $\varphi^N$ in the left-hand side of Eq.~\eqref{eq:StartingPointDem} and in the right-hand side of Eq.~\eqref{eq:EndPointDem} demonstrates Eq.~\eqref{eq:TheMainResult}. 
In order to prove \eqref{eq:TheMainResult} in the grand canonical case, it suffices to replace $\Xi(\varphi)$ by 1 in the left-hand sides of \eqref{eq:MiddlePointDem} and \eqref{eq:EndPointDem} and use the absence of correlation between occupation numbers Eq.~\eqref{eq:TrivialUncorrelationsGC}.
\end{widetext}  

\subsubsection{Generalization}

The same technique allows us to generalize Eq.~\eqref{eq:TheMainResult} straightforwardly~: the $p$-point correlation function can be expressed in terms of $q$-point correlation functions for any $q<p$. For instance for $q=2$ we have
\begin{equation}
  \label{eq:geneq2}
    \overline{n_1n_2\cdots n_p} 
    = 
  \mp
    \frac{ 
    \left|\begin{array}{cc}
     \overline{n_1n_3\ldots n_p} & x_1  \\
     \overline{n_2n_3\ldots n_p} & x_2    \end{array}\right|
    }{
    \left|\begin{array}{cc}
     1 & x_1  \\
     1 & x_2  
    \end{array}\right|
    },
\end{equation}
and other such relations obtained by picking different indices among the $p(p-1)/2$ pairs of indices (such an expression was obtained for fermions in Ref.~\cite{Sch17}). 
More generally, as a function of $q$-point correlation functions we have
\begin{equation}
  \label{eq:geneqarb}
    \overline{n_1n_2\cdots n_p} 
    = 
  (\mp)^{q-1}
    \frac{ 
    \left|\begin{array}{cccc}
     \overline{n_1n_{q+1}\ldots n_p} & x_1 & \cdots & x_1^{q-1} \\
     \overline{n_2n_{q+1}\ldots n_p} & x_2 & \cdots & x_2^{q-1} \\
     \vdots & \vdots & \ddots & \vdots \\
     \overline{n_qn_{q+1}\ldots n_p} & x_q & \cdots & x_q^{q-1}
    \end{array}\right|
    }{
    \left|\begin{array}{cccc}
     1 & x_1 & \cdots & x_1^{q-1} \\
     1 & x_2 & \cdots & x_2^{q-1} \\
     \vdots & \vdots & \ddots & \vdots \\
     1 & x_q & \cdots & x_q^{q-1}
    \end{array}\right|
    }
\end{equation}
and other such relations obtained by picking any $q$ indices among the $\binom{p}{q}$ possibilities. The steps are exactly the same as in \eqref{eq:StartingPointDem}--\eqref{eq:EndPointDem}, replacing the correlators in the determinant by their expression \eqref{eq:BeautifulGeneratingFunction}. For $q=p$ we obviously recover \eqref{eq:TheMainResult}. Again, substituting $\Xi(\varphi)$ by 1 in the proof shows that Eq.~\eqref{eq:geneqarb} is valid also for the grand canonical ensemble.

\subsubsection{Relation with Schur polynomials}

The ratio of determinants in Eq.~\eqref{eq:TheMainResult} allows us to identify another interesting connection with the theory of symmetric functions, namely with a fourth family of symmetric functions, the so-called Schur polynomials \cite{Mac95}. The definition of these polynomials is given in Appendix~\ref{app:Schur}; they are expressed as a ratio of two determinants.

The Schur polynomials naturally appear if one replaces the $\overline{n_i}$ in the numerator of Eq.~\eqref{eq:TheMainResult} by their expression \eqref{eq:BEandFDcanonical}. By linear expansion of the determinant with respect to its first column we directly obtain
\begin{align}
\label{eq:withschur}
\overline{n_1\cdots n_p}
=  
\frac{x_1\cdots x_p}{Z_N}
\sum_{k=p}^N (\pm1)^{p+k}\,Z_{N-k}\,
s_{\varpi_k}(x_1,\cdots,x_p)
\:,
\end{align}
where $s_{\varpi_k}$ are the Schur polynomials for the partition $\varpi_k=(k-p,0^{p-1})$, cf. Appendix~\ref{app:Schur}.
We recall that $Z_{N-k}$ is identified with either the elementary symmetric polynomial $\es_{N-k}(x_1,\ldots,x_M)$ for fermions (lower sign $-$), or the homogeneous symmetric polynomial $\hs_{N-k}(x_1,\ldots,x_M)$ for bosons (upper sign $+$), where $M$ is the dimension of the one-particle Hilbert space.
Introducing the Schur functions has allowed us to reduce the $p$-fold sum in Eq.~\eqref{eq:TheMainResult0} to a single sum in \eqref{eq:withschur}.
The relation between the two expressions relies on the representation
\begin{align}
     &s_{(k-p,0^{p-1})}(x_1,\cdots,x_p)
     \nonumber\\
     = &\sum_{k_1=1}^N \cdots\sum_{k_p=1}^N
     \delta_{k,k_1+\cdots+k_p}\, 
     x_1^{k_1-1}\cdots x_p^{k_p-1}
     \:.
\end{align} 

Let us illustrate the formula~\eqref{eq:withschur}.
The case $p=1$ is rather trivial, as $\varpi_k=(k-1)$ and thus $x_1\,s_{(k-1)}(x_1)=x_1^k$, making the correspondence with \eqref{eq:BEandFDcanonical} obvious.
The case $p=2$ is more interesting~: the partitions are $\varpi_k=(k-2,0)$, from which we write (cf. Appendix~\ref{app:Schur}) 
\begin{align}
   x_1x_2 &\,s_{(k-2,0)}(x_1,x_2) 
   =  x_1x_2\,  
   \frac{
      \left|\begin{array}{cc} x_1^{k-1} & 1 \\[0.125cm] x_2^{k-1} & 1 \end{array}\right|
      }{\left|\begin{array}{cc} x_1 & 1 \\ x_2 & 1 \end{array}\right|}
      \nonumber\\
  & =  x_1^{k-1}x_2+x_1^{k-2}x_2^2+\cdots+x_1x_2^{k-1} 
  \:. 
  \end{align}

\subsection{Correlation functions with higher powers (bosons)~: proof of Eq.~\eqref{eq:GeneralCorrelations}}

In the case of bosons, we can also consider higher moments of the occupation numbers (for fermions we have of course $n_\lambda^r=n_\lambda$). This question has attracted a lot of attention for the characterisation of the number of condensed bosons in a BEC \cite{Pol96,GroHol96,GajRza97,WilWei97,HolKalKir98}. In the grand canonical ensemble, the integer moments of each occupation number can be obtained simply from the individual grand partition function $\xi_\lambda=(1-\varphi x_\lambda)^{-1}$ for individual eigenstate $\ket{\lambda}$ as 
\begin{align}
  \overline{n_\lambda^r}^\gc
  = \frac{1}{\xi_\lambda} \left(\varphi\deriv{}{\varphi}\right)^r \xi_\lambda
  = \sum_{k=1}^\infty  a_k(r)\, (x_\lambda\varphi)^k
  \:,
\end{align}
with $ a_k(r)=k^r-(k-1)^r $ and $r\in\mathbb{N}$.
Applying \eqref{eq:GCandCANmeans} to $\mathcal{A}(\{n_\lambda\})=n_1^{r_1}\cdots n_p^{r_p}$, and making use of the independence of the occupation numbers in the grand canonical ensemble as in \eqref{eq:TrivialUncorrelationsGC}, we readily obtain \eqref{eq:GeneralCorrelations}. For instance for the second moment we have
\begin{equation}
    \label{eq:2momOccupation}
    \overline{n_\lambda^2} 
  = \sum_{k=1}^N (2k-1) \frac{Z_{N-k}}{Z_N} \, x_\lambda^{k}
  \:.
\end{equation}
This representation will be of practical use in the following section.


\section{Condensation of bosons in a 1D harmonic trap}
\label{sec:BEC1D}

As a simple illustration of our results, we consider $N$ bosons in a 1D harmonic trap with frequency $\omega$. 
The problem has been studied within the grand canonical~\cite{BroDevLem96,KetDru96,Mul97,PetSmi02,PetGanShl04}, the canonical~\cite{BroDevLem96,WilWei97,WeiWil97,BorHarMulHil99} and the microcanonical~\cite{Tem49,GroHol96,GajRza97,WeiWil97} ensembles~\cite{footnote1}. 
In particular some limiting behaviors of the ground state occupancy for $T\to0$, recalled below, were obtained in several of these references.
The probability distribution of the occupation number of the $k$-th level was obtained in Ref.~\cite{WilWei97}~:
\begin{equation}
  \label{eq:WilkensWeis1997}
  \mathscr{P}_{k,N}(n) = \overline{\delta_{n_k,n}}^{(N)}
  = x^{kn}\left[ 1- x^k + x^{N-n+k}\right] \frac{Z_{N-n}}{Z_{N}}
  \:,
\end{equation}
where $x=\EXP{-\beta\omega}$.
It is however not straightforward to extract simple information, like moments or cumulants, or to analyze the large-$N$ asymptotics of this distribution.
Here we show that the canonical formulae obtained in the previous sections lead to simple analytical results appropriate to discuss the large $N$ limit.

\subsection{Thermodynamic properties}

Up to a shift in energy, the one-body spectrum is $\varepsilon_n=n\,\omega$ for $n\in\mathbb{N}$ (we set $\hbar=\kB=1$).
The $N$-body partition function \cite{TodKubSai92,TexRou17book}
\begin{equation}
  \label{eq:ZNBosons}
  Z_N=\prod_{n=1}^N(1-\EXP{-n\beta\omega})^{-1}
\end{equation}
corresponds to $N$ independent bosonic modes with frequencies $\Omega_n=n\,\omega$ for $n=1,\ldots,N$.

The problem involves three characteristic temperature scales. 
(\textit{i})
The lowest scale, $T_Q=\omega$, separates the regime where the spectrum should be considered discrete ($T\ll T_Q$) from the one where it can be described as a continuous spectrum ($T\gg T_Q$).
(\textit{ii})
The scale $T_*=N\omega$ separates the quantum regime $T\ll T_*$, where the upper modes are frozen in their ground state (see Eq.~\eqref{eq:ZNBosons}), from the classical regime  $T\gg T_*$ where all the modes can be described as classical oscillators, in which case we recover the Maxwell-Boltzmann partition function $Z_N\simeq(1/N!)(\beta\omega)^{-N}$ corresponding to neglecting the effect of quantum correlations.
(\textit{iii})
The third temperature scale is the Bose-Einstein temperature 
\begin{equation}
  \TB = \frac{N\omega}{\ln N},
\end{equation}
below which a macroscopic fraction of bosons accumulates in the individual ground state.
It can be obtained from the analysis of the canonical chemical potential $\mu^\can=F_{N}-F_{N-1}$ or the fugacity $\varphi^\can=\EXP{\beta\mu^\can}=Z_{N-1}/Z_N=1-\EXP{-N\beta\omega}$.
Introducing (incorrectly) this expression in the grand canonical expression of the ground state occupancy, $\overline{n_0}^\gc=\big[1/\varphi-1\big]^{-1}$, Eq.~\eqref{eq:BEandFD}, shows that $\overline{n_0}^\gc\sim N$ for $\varphi\sim1-1/N$ i.e. $T\sim \TB=N\omega/\ln N$.


In the following, we will not describe the effect of the discrete nature of the spectrum, and will always consider the limit $T\gg\omega$ (the condition will be implicit in the rest of the paper).
In particular, we can treat the sum over the spectrum in $\ln Z_N$ as an integral, so that \eqref{eq:ZNBosons} yields 
$\ln Z_N\simeq-\int_0^N\D n\,\ln(1-\EXP{-n\beta\omega})$. 
The latter expression can be reformulated in terms of the polylogarithm function 
$\mathrm{Li}_2(x)=\sum_{n=1}^\infty x^n/n^2=-\int_{-\ln x}^{+\infty}\D y\,\ln(1-\EXP{-y})$ (see \S25~of \cite{DLMF}). 
The free energy is then given by
\begin{equation}
  \label{eq:FNBosons}
  F_N=-\frac{1}{\beta}\ln Z_N
  \simeq
  \frac{ \mathrm{Li}_2(\EXP{-N\beta\omega})-\mathrm{Li}_2(1) }{\beta^2\omega}
  \:.
\end{equation}

\subsection{Mean occupation numbers}

\subsubsection{Classical regime: $T\gg T_*$}

In the classical regime $T\gg T_*$, i.e.~$N\beta\omega\ll 1$, one can use the asymptotic behavior $\mathrm{Li}_2(1-\epsilon)\simeq\pi^2/6+\epsilon\,(\ln\epsilon-1)$ for $\epsilon\ll1$. From Eq.~\eqref{eq:FNBosons} we recover the classical (Maxwell-Boltzmann) result $Z_N\sim\EXP{N}\,(N\beta\omega)^{-N}$, which coincides with the expression given above since $N!\sim N^N\EXP{-N}$. The mean occupation number, given by Eq.~\eqref{eq:BEandFDcanonical}, is dominated by the first term, so that for the $k$-th level $\varepsilon_k$ it is given by $\overline{n_k} \simeq N\beta\omega\,\EXP{-\beta\varepsilon_k}$. Since the canonical fugacity behaves as $\varphi^\can\simeq N\beta\omega$, we get $\overline{n_k}\simeq\varphi^\can\,\EXP{-\beta\varepsilon_k}$, which coincides with the well-known grand canonical behavior.

\subsubsection{Quantum regime: $T\ll T_*$}

We now turn to the more interesting regime where $T\ll T_*$ (and of course $T\gg T_Q$). 
The sum in \eqref{eq:BEandFDcanonical} can be replaced by an integral. 
Using \eqref{eq:FNBosons} for the expression of $Z_N$, the mean occupation number can be reexpressed as
\begin{align}
  \label{eq:NKintegral}
     &\overline{n_k} 
     \simeq
     N \int_0^1\D y\, 
     \\\nonumber
     &\times
     \exp
     \left\{
       -N\beta\varepsilon_k y
       +\frac{\mathrm{Li}_2(\EXP{-N\beta\omega}) - \mathrm{Li}_2(\EXP{-N\beta\omega(1-y)}) }{\beta\omega} 
     \right\}
     \:,
\end{align}
where $\varepsilon_k=k\omega$.
While the exact form \eqref{eq:BEandFDcanonical} is only tractable for small $N$ in pratice, the integral representation \eqref{eq:NKintegral} has the advantage that it allows to study the occupation without restriction on $N$. One must however keep in mind that \eqref{eq:NKintegral} only holds in the intermediate regime $\omega\ll T\ll N\omega$. 


The integral expression \eqref{eq:NKintegral} can be further simplified using the behavior of the polylogarithm function in the vicinity of 0, $\mathrm{Li}_2(x)\simeq x$ for $x\to0$. Indeed, in the regime where $N\beta\omega\gg 1$, one can replace $\mathrm{Li}_2(\EXP{-N\beta\omega})$ by $\EXP{-N\beta\omega}$. Moreover, when $\beta\omega\ll 1$ one can also replace $\mathrm{Li}_2(\EXP{-N\beta\omega(1-y)})$ by $\EXP{-N\beta\omega(1-y)}$, since in the vicinity of $y=1$, where this approximation breaks down, the integrand becomes proportional to $\EXP{-1/(\beta\omega)}\ll 1$. For the same reason one can extend the integral to infinity. Equation \eqref{eq:NKintegral} thus reduces to
\begin{equation}
  \label{eq:NKintegral2}
  \overline{n_k} 
     \simeq
     N \int_0^{\infty}\D y\,
     \exp
     \left\{
       -N\beta\varepsilon_k y
       +\frac{\EXP{-N\beta\omega}-\EXP{-N\beta\omega(1-y)} }{\beta\omega} 
     \right\}
     \:.
\end{equation}
Introducing the parameter 
\begin{equation} 
  \label{eq:DefZ}
  z = \frac{\EXP{-N\beta\omega}}{\beta\omega}
  = \frac{T}{\omega}\,N^{-\TB/T}
\end{equation}
and making the change of variables $u=z\EXP{N\beta\omega y}$, Eq.~\eqref{eq:NKintegral2} yields a representation in terms of the incomplete Gamma function,
\begin{equation}
  \label{eq:MeanNk}
    \overline{n_k} 
     \simeq \frac{1}{\beta\omega}\,
     z^k\EXP{z}\, \Gamma(-k,z)
     \:.
\end{equation}
In the inset of Fig.~\ref{fig:1Dcondensate} we compare \eqref{eq:MeanNk} with the exact expression \eqref{eq:BEandFDcanonical}~: the difference is below $1\%$ on the interval $[0,\TB]$ for relatively small $N$ and we can see that the temperature range over which the difference remains small grows as $N$ increases.

\subsubsection{Ground state}

\begin{figure}[!ht]
\centering
\includegraphics[width=0.45\textwidth]{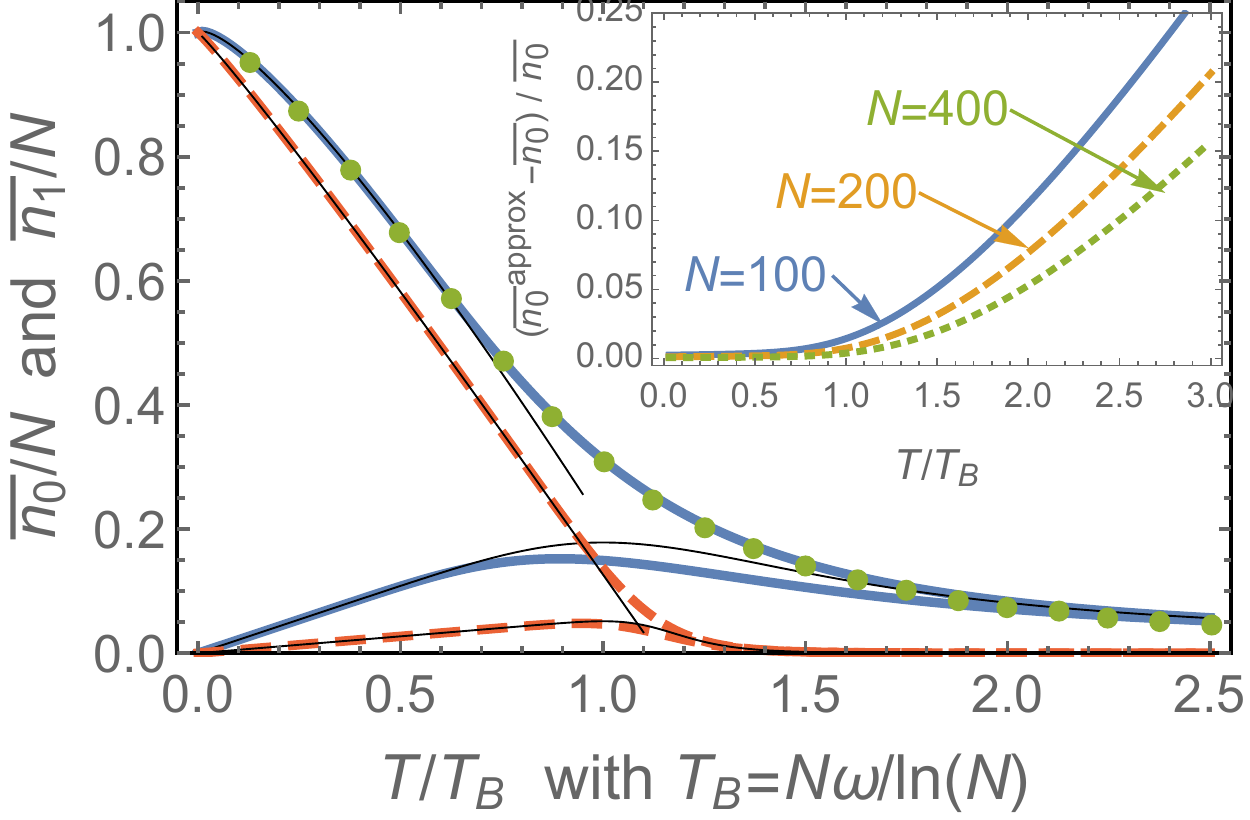}
\caption{(Color online)
{\it Occupations of the ground state and the first excited state, Eq.~\eqref{eq:MeanNk}, for $N=100$ (blue solid line) and $N=10^8$ (red dashed). 
The green dots correspond to the exact result \eqref{eq:BEandFDcanonical} for the ground state with $N=100$.
Thin black lines are \eqref{eq:pseudoBEC} and \eqref{eq:n1approx}.}
Inset~:
\textit{Relative difference between the approximate form \eqref{eq:MeanNk} and the exact form \eqref{eq:BEandFDcanonical}.} 
}
\label{fig:1Dcondensate}
\end{figure}

Setting $k=0$ in Eq.~\eqref{eq:MeanNk}, the incomplete Gamma function reduces to the exponential integral $\Gamma(0,z)=E_1(z)$. 
The $T\ll\TB$ regime corresponds to the limiting behavior $E_1(z)=\ln[\EXP{-\gamma}/z]+\mathcal{O}(z)$ for $z\to0$, where $\gamma\simeq0.577$ is the Euler-Mascheroni constant. 
Hence 
\begin{equation}
  \label{eq:pseudoBEC}
  \frac{\overline{n_0}}{N}\simeq 1-\frac{T\ln(\EXP{\gamma}T/\omega)}{N\omega}
  \hspace{0.5cm}\mbox{for }
  T\ll \TB
  \:.
\end{equation}
This behavior was already obtained in Ref.~\cite{HolKalKir98} by a different approach
(see also~Appendix~\ref{app:GC}, where we recall how the limiting behavior can be obtained within a grand canonical treatment \cite{PetGanShl04}). In Fig.~\ref{fig:1Dcondensate} we compare the approximate expression \eqref{eq:pseudoBEC} with the exact sum \eqref{eq:BEandFDcanonical}. For large enough $N$ the behavior \eqref{eq:pseudoBEC} is indistinguishable from the exact result up to $T\sim\TB$.

\subsubsection{Excited states}

For the excited states, because of the factor $\EXP{-N\beta\varepsilon_k y}$, the integral \eqref{eq:NKintegral} is dominated by the neighbourhood of the lower boundary. In this case we can use $\EXP{-N\beta\omega}-\EXP{-N\beta\omega(1-y)}\simeq -N\beta\omega y \EXP{-N\beta\omega}$ for $y\to0$.
This is equivalent to replacing \eqref{eq:MeanNk} by the approximation $z^k\EXP{z}\, \Gamma(-k,z)\simeq1/(k+z)$, which corresponds to the interpolation between the two limiting behaviors 
$\simeq1/k$ for $z\to0$ and $\simeq1/z$ for $z\to\infty$ (the agreement with $1/(k+z)$ becomes excellent at large $k$). 

As a result we get the approximate form
\begin{equation}
  \label{eq:n1approx}
  \overline{n_k} \simeq \frac{T}{\varepsilon_k+T\,\EXP{-N\omega/T}}
\end{equation}
for $k\geq1$. 
For $T\ll\TB$ we get the linear behavior $\overline{n_k}\simeq T/\varepsilon_k$. 
However, unlike \eqref{eq:pseudoBEC}, Eq.~\eqref{eq:n1approx} also describes the crossover from this linear behavior to the decaying behavior above $\TB$, as illustrated in Fig.~\ref{fig:1Dcondensate}. 
Equation~\eqref{eq:n1approx} shows that the mean occupation reaches its maximum for $T=N\omega/\ln(N/k)\simeq \TB$, with 
\begin{equation}
  \overline{n_k}\big|_\mathrm{max} \simeq \frac{N/k}{\ln(N/k)}
  \:.
\end{equation}
The presence of the logarithm in the denominator shows that only the ground state has a macroscopic occupation below $\TB$, as expected when BEC occurs. 

For the highest excited states, such that $\varepsilon_k\gg T$, the continuous approximation of the sum \eqref{eq:BEandFDcanonical}, leading to the integral \eqref{eq:NKintegral}, fails, and the occupation decays exponentially as $\overline{n_k}\simeq\EXP{-\beta\varepsilon_k}$.
This is the expected classical behavior (for $\varphi^\can\simeq1$), as weakly occupied levels can be considered in the classical regime.

\subsection{Variance of the ground state occupation}

\begin{figure}[!ht]
\centering
\includegraphics[width=0.45\textwidth]{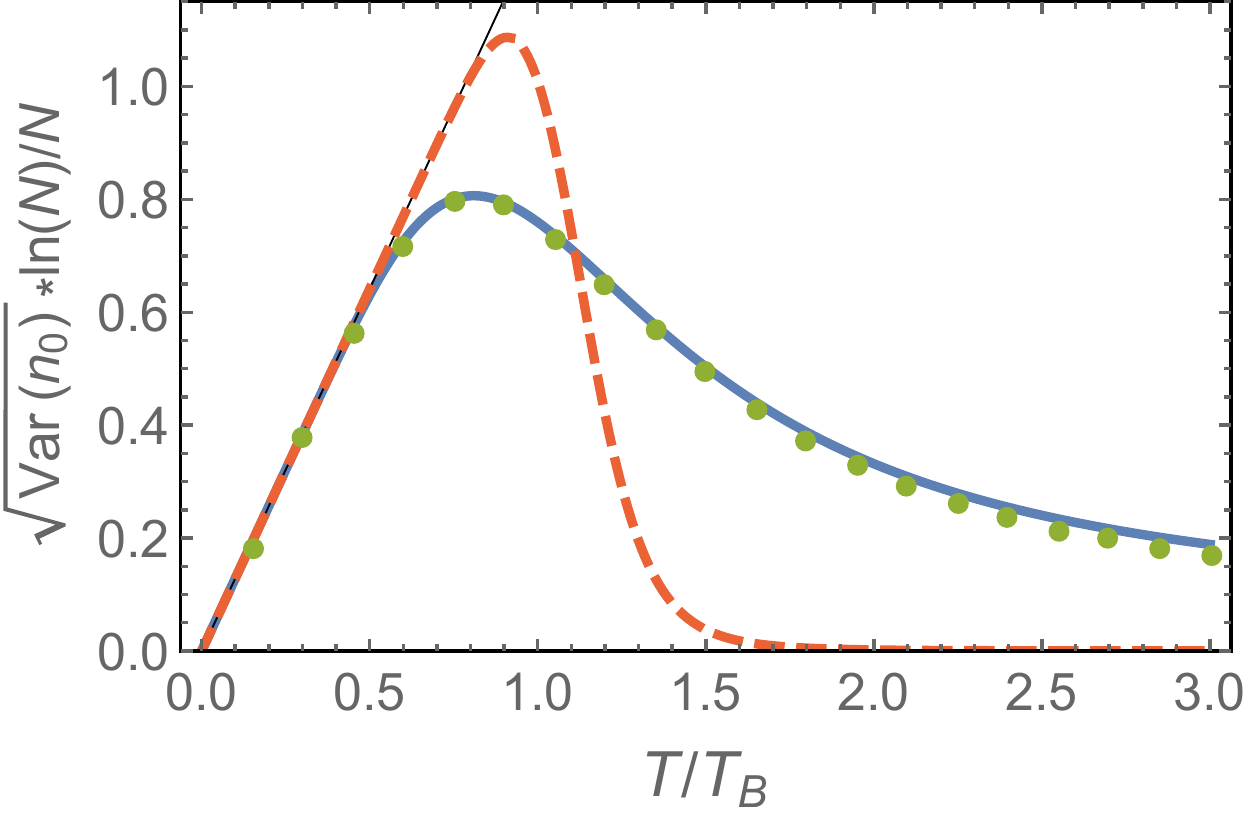}
\caption{(Color online) \textit{Fluctuations of the ground state occupancy as a function of the rescaled temperature, for $N=100$ (blue solid line) and $10^8$ (red dashed), from Eq.~\eqref{eq:MomRNk}. 
Thin black line corresponds to \eqref{eq:VarN0lowT}.
Green dots~: exact expression from \eqref{eq:BEandFDcanonical}-\eqref{eq:2momOccupation}.}
}
\label{fig:1DcondensateVar}
\end{figure}

We now study the fluctuations around the mean occupation number. We restrict ourselves to the ground state, which has attracted some attention in higher dimension \cite{Pol96,HolKalKir98}. The exact expression for $\overline{n_0^2}$ is given by \eqref{eq:2momOccupation}. In the most interesting regime, $T_Q=\omega\ll T\ll T_*=N\omega$, we get an integral representation similar to \eqref{eq:NKintegral}~: the coefficient $a_k(2)=2k-1$ in the sum \eqref{eq:2momOccupation} translates into a factor $2Ny-1\simeq2Ny$ in the integral \eqref{eq:NKintegral}, so that 
\begin{align}
  \label{eq:VarN0integral}
     \overline{n_0^2} 
     \simeq
     2N^2 \int_0^1\D y\,y\, 
     \EXP{[ \mathrm{Li}_2(\EXP{-N\beta\omega}) - \mathrm{Li}_2(\EXP{-N\beta\omega(1-y)}) ]/(\beta\omega)}
     \:.
\end{align}
Performing the same approximations as above and using the asymptotics
$\int_x^\infty\D u\,\ln(u)\,\EXP{-u}/u\simeq(1/2)\big[-\ln^2x+\gamma^2+\pi^2/6\big]$ as $x\to0$, we obtain
\begin{equation}
  \label{eq:VarN0lowT}
  \mathrm{Var}(n_0)\simeq \frac{1}{6} \left(\frac{\pi T}{\omega}\right)^2
  \hspace{0.5cm}\mbox{for }
  T\ll \TB
  \:.
\end{equation}
This behavior was obtained in Refs.~\cite{WilWei97,HolKalKir98} by a different canonical 
calculation~\cite{footnote2}. 
It was also obtained within the microcanonical ensemble in Refs.~\cite{GroHol96,GajRza97}.
The variance deduced from \eqref{eq:VarN0integral},
which is plotted in Fig.~\ref{fig:1DcondensateVar}, presents a peak close to $T\sim \TB$, with a scaling $\mathrm{Var}(n_0)|_{T=\TB}\sim N^2/\ln^2N$. 
A careful analysis of the expression \eqref{eq:MomRNk} derived below, with the help of the software \texttt{Mathematica},  
shows that the peak in the variance scales as
\begin{equation}
   \mathrm{Var}(n_0)\big|_\mathrm{max}
   \simeq 
   \frac{\pi^2 N^2}{6\ln^2N}
   \left[
      1 -  \frac{(\ln\ln N)^2}{\ln N} + \mathcal{O}\left(\frac{\ln\ln N}{\ln N}\right)
   \right],
\end{equation}
thus the relative fluctuations are $\delta n_0/\overline{n_0}\sim1/\ln N$.
Due to this slow decay, $n_0$ cannot be considered self-averaging in practice.

\subsection{Distribution of the occupation numbers}

\subsubsection{Characteristic function}
In order to demonstrate the efficiency of our formalism, we now derive the full distribution of the ground state occupancy.
We start from the general expression of the moments, Eq.~\eqref{eq:GeneralCorrelations}. 
As for the first two moments, we replace the sum by an integral, which is valid for $T_Q=\omega\ll T\ll T_*=N\omega$. 
For $r\geq 1$, we have
\begin{align}
     \overline{n_k^r} 
     \simeq&
     N^{r+1} \int_0^1\D y\,
     \left[ y^{r} - \left(y-\frac1N\right)^r\right]\,\EXP{-N\beta\varepsilon_k y} 
     \nonumber
     \\
     &\times
     \exp\left\{ 
        \frac{ \mathrm{Li}_2(\EXP{-N\beta\omega}) - \mathrm{Li}_2(\EXP{-N\beta\omega(1-y)}) }{\beta\omega}
         \right\}
     \:,
\end{align}
where $\varepsilon_k=k\omega$.
Weighting this expression by $(\alpha\beta\omega)^r/r!$ and summing over $r$, we get the characteristic function
\begin{align}
  \label{eq:RepresGF1}
  &\overline{ \exp\{\alpha\,\beta\omega\,n_k\} } 
  \simeq 
  1 + \frac{1-\EXP{-\alpha\,\beta\omega}}{\beta\omega}\,z^{k-\alpha}
     \\
     \nonumber
     &\times
    \int_{z}^{1/(\beta\omega)}\D u\, u^{\alpha-k-1}\,
     \exp\left\{ 
        \frac{ \mathrm{Li}_2(\beta\omega z) - \mathrm{Li}_2(\beta\omega u) }{\beta\omega}
         \right\}
         \:,
\end{align}
where $z$ is given by \eqref{eq:DefZ} and we made the change of variables $u=z\EXP{N\beta\omega y}$. In the limit where $\beta\omega\to0$ (continous approximation), keeping $z$ finite (i.e. probing $T\sim\TB$), one can again use $\mathrm{Li}_2(x)\simeq x$ for $x\to0$. 
We recognize the integral representation of the incomplete Gamma function $\Gamma(\alpha-k,z)$.
We finally get
\begin{equation}
  \label{eq:CGFNk}
   \overline{ \EXP{\alpha\,\beta\omega\,n_k} }
   \simeq 
    1+\alpha\,\EXP{z}\,z^{k-\alpha} \,\Gamma(\alpha-k,z)
    \equiv \Gf_{k,z}(\alpha) 
    \:,
 \end{equation}
where we introduced the moment generating function $\Gf_{k,z}(\alpha)$. We define a rescaled variable $\xi$ by $\Gf_{k,z}(\alpha)=\overline{\EXP{\alpha\xi}}$. It is such that $\xi\simeq\beta\omega\,n_k$.
In this regime, the generating function of the occupation number for the $k$-th excited state only depends on the non-trivial combination of parameters given by $z$ defined in Eq.~\eqref{eq:DefZ}.

We can get the moments by expanding the generating function as
\begin{equation}
  \Gf_{k,z}(\alpha) =\sum_{r=0}^\infty \frac{\mu_r(z,k)}{r!}\alpha^r
  \hspace{0.5cm}\mbox{with }
  \overline{n_k^r} \simeq \frac{\mu_r(z,k)}{(\beta\omega)^r}
  \:.
\end{equation}
We get
\begin{equation}
  \label{eq:Mom1Nk}
  \mu_1(z,k) = z^k\EXP{z}\, \Gamma(-k,z) 
  \:,
\end{equation}
which coincides with Eq.~\eqref{eq:MeanNk}, as it should.
The higher moments are expressed in terms of Meijer-$G$ functions~:
\begin{align}
  \label{eq:MomRNk}
  \mu_r(z,k)&=  r!\, \EXP{z}\, G_{r,r+1}^{r+1,0}
  \left(
     z 
     \left| 
     \begin{array}{c} 1+k,\ldots, 1+k  \\ 0, k,\ldots, k  \end{array}
     \right.
  \right)
  \:.
\end{align}
These expressions are valid in the full range of temperature $T\sim\TB$ where quantum correlations dominate. 
The cumulants of the occupations can be deduced from the expansion
\begin{equation}
  \label{eq:48}
  \ln\Gf_{k,z}(\alpha)
  =\sum_{r=1}^\infty \frac{\cum_r(z,k)}{r!}\alpha^r
  \hspace{0.5cm}\mbox{with }
  \overline{n_k^r}^\mathrm{cum} \simeq \frac{\cum_r(z,k)}{(\beta\omega)^r}
  \:.
\end{equation}
However we have not found any expression for the cumulants that would be simpler than that obtained from the moments.

\subsubsection{Distribution}
\label{subsubsec:DistributionPk}

Using the integral representation 
$\Gamma(a,z)=z^a\int_0^\infty\D\xi\, \exp\big\{a\,\xi-z\EXP{\xi}\big\}$, 
we rewrite the generating function \eqref{eq:CGFNk} as 
\begin{equation}
  \Gf_{k,z}(\alpha) = \overline{\EXP{\alpha\xi}}
  =1 + \EXP{z}\int_0^\infty\D\xi\, \alpha\, \EXP{\alpha\xi}\,\EXP{-k\xi-z\EXP{\xi}}
  \:.
\end{equation}
An integration by parts makes it clear that the rescaled occupation number $\xi\simeq\beta\omega\,n_k$ is distributed according to the law 
\begin{equation}
  \label{eq:DistributionNk}
  Q_{k,z}(\xi) = \heaviside(\xi)\,\EXP{z}
  \left(k+z\EXP{\xi}\right)\,\exp\left\{-k\xi-z\EXP{\xi}\right\}
  \:,
\end{equation}
where $\heaviside(\xi)$ is the Heaviside function.
Although the connection is not obvious, this distribution is the large $N$ limit of \eqref{eq:WilkensWeis1997}.

We can compare our result \eqref{eq:DistributionNk} with the simple result given by the grand canonical ensemble. In this case, occupations are independent and the distribution of the occupation is exponential, $\mathscr{P}_{k}^\gc(n)\propto\varphi^n\EXP{-n\beta\varepsilon_k}$, where $\varphi$ is the fugacity, cf. Eq.~\eqref{eq:DistribGrandCano}.
The rescaled variable $\xi\simeq\beta\omega\,n_k$ is then distributed according to the law $Q_{k}^\gc(\xi)\propto\varphi^{T\xi/\omega}\,\EXP{-k\xi}$.
To make the correspondence more clear, we replace $\varphi$ by the canonical fugacity $\varphi^\can=1-\EXP{-N\beta\omega}$~; we get the form 
$Q_{k}^\gc(\xi)\propto\EXP{-(z+k)\xi}$.
The two distributions thus significantly differ, and in particular the large deviations, as shown in the inset of Fig.~\ref{fig:DitribN1}.

As stressed by Sch\"onhammer~\cite{Sch17}, the deviation from the purely exponential distribution in the canonical ensemble can be interpreted as a deviation from Wick theorem induced by the constraint on the number of particle number.

\subsubsection{Ground state} 

\begin{figure}[!ht]
\centering
\includegraphics[width=0.4\textwidth]{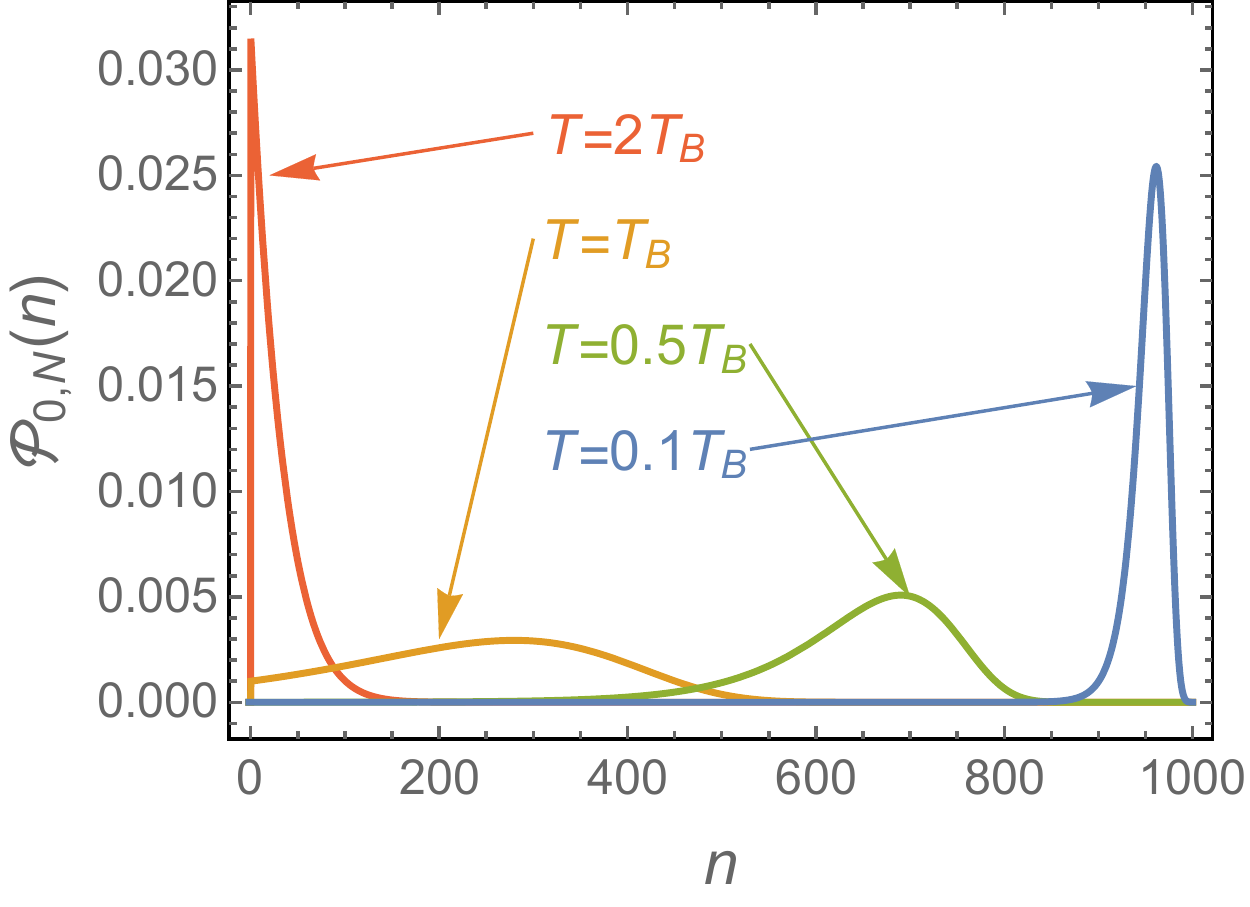}
\caption{(Color online) \textit{ Distribution $\mathscr{P}_{0,N}(n)$ of the number of condensed bosons for $N=1000$. Temperature is (from left to right) $T/\TB=2$ (red), $1$ (orange), $0.5$ (green) and $0.1$ (blue), obtained from Eq.~\eqref{eq:DistribN0}.}
}
\label{fig:DitribN0}
\end{figure}

In the case of the ground state, it is more convenient to shift the rescaled variable as 
$\zeta=\xi+\ln z\simeq\beta\omega\,n_0+\ln z$.
The new variable is thus distributed according to 
\begin{equation}
\label{fz}
  F_z(\zeta)\equiv Q_{0,z}(\xi) = \heaviside(\zeta-\ln z)\, 
  \exp\left\{z+\zeta - \EXP{\zeta} \right\}
  \:,
\end{equation}
which is the \textit{truncated} Gumbel distribution.
In the limit $z\to0$, we get the Gumbel law $F_0(\zeta)=\exp\left\{\zeta - \EXP{\zeta} \right\}$, defined on $\mathbb{R}$, describing extreme value statistics of independent random variables~\cite{Gum35,Gum58}. 
The probability distribution for the ground state occupancy $n_0$ can then be written as
\begin{equation}
  \label{eq:DistribN0}
  \mathscr{P}_{0,N}(n)
  \simeq
  \frac{\omega}{T}\,
  F_z\!\left( \frac{\omega}{T}\left(n-N+ \frac{T}{\omega}\ln(T/\omega)\right) \right)
  \:.
\end{equation}
This distribution is plotted in Fig.~\ref{fig:DitribN0} for different temperatures. 
The curves correspond at first sight with the plot of \eqref{eq:WilkensWeis1997} in \cite{WilWei97}, although the connection with the Gumbel distribution was not made in that paper.

The distribution simplifies in the regime $T\ll\TB$, as well as the moments: 
when $z\to0$, Eq.~\eqref{eq:CGFNk} yields $\ln\Gf_{0,z}(\alpha) \simeq -\alpha\,\ln z + \ln \Gamma(1+\alpha)$, leading to 
\begin{align}
  \ln\Gf_{0,z}(\alpha)
  \simeq -\alpha\,\ln z  + \sum_{r=1}^\infty \frac{\psi^{(r-1)}(1)}{r!}\,\alpha^r
  \hspace{0.5cm}\mbox{as }
  z\to0
  \:,
\end{align}
where $\psi(x)$ is the digamma function. This leads in particular to
$\mu_1(z,0)=\cum_1(z,0)=-\ln z+\psi(1)+\mathcal{O}(z)\simeq-\ln z-\gamma$, 
in accordance with \eqref{eq:pseudoBEC}, and 
$\cum_2(z,0)=\psi'(1)+\mathcal{O}(z\ln z)\simeq\pi^2/6$, 
in accordance with \eqref{eq:VarN0lowT}.   
In general, we have for $r\geq 2$ the expression $\cum_r(z,0)=\psi^{(r-1)}(1)+\mathcal{O}(z\ln^{r-1} z)$ as $z\to0$, thus 
\begin{equation}
  \label{eq:CumulantN0lowT}
  \overline{n_0^r}^\mathrm{cum} \simeq \psi^{(r-1)}(1)\,\left(\frac{T}{\omega}\right)^r
  \hspace{0.5cm}\mbox{for } r\geq 2
  \:,
\end{equation}
which coincide with the cumulants of the Gumbel law, as it should. 
Again, recall that this behavior holds in the regime $T_Q\ll T\ll \TB$.

\subsubsection{Excited states}
\label{subsubsec:ExcitedStates}

The study of the fluctuations of the occupation numbers for the excited states follows the same lines as for the ground state. For instance the second moment $\overline{n_k^2}$ is given by inserting a factor $2N\,y$ in the integral \eqref{eq:NKintegral}. Similar approximations as for the calculation of the mean value lead to $\overline{n_k^2}\simeq2\left( \overline{n_k} \right)^2$, thus 
\begin{equation}
  \mathrm{Var}(n_k) \simeq \left( \overline{n_k} \right)^2
  \hspace{0.5cm}\mbox{for } 
  T \ll  \TB
  \:.
\end{equation}
As is turns out, this approximation reproduces quite well the variance in the whole regime $T \ll  T_*$.
As for the ground state, we get a quadratic behavior at low temperature,
$\mathrm{Var}(n_k)\sim(T/\varepsilon_k)^2$ for $\omega\ll T\ll \TB$.
The fluctuations are maximum for $T \simeq \TB$ with $\mathrm{Var}(n_k)\big|_\mathrm{max}\simeq (N/k)^2/\ln^2(N/k)$.
Hence, the maximal
fluctuations in the excited states are of the same order as the fluctuations in the ground state
\begin{equation}
  \delta n_k \sim 
  \frac1k\, \delta n_0 \hspace{0.5cm}\mbox{for } T\sim \TB
  \:,
\end{equation}
however the relative fluctuations are larger in the excited states, $\delta n_k/\overline{n_k}\sim1$, than in the ground state~$\delta n_0/\overline{n_0}\sim1/\ln N$.

The distribution of the occupation number is given by \eqref{eq:DistributionNk}.
We remark that the distribution simplifies in the low temperature limit as $Q_{k,0}(\xi)=\heaviside(\xi)\,k\,\EXP{-k\xi}$, i.e.
$
  \mathscr{P}_{k,N}(n)
  \simeq 
  \beta\varepsilon_k\,\EXP{-n\beta\varepsilon_k}
$
for 
$ T \ll  \TB$.
Interestingly, in this limit, this distribution coincides with the similar distribution obtained in the grand canonical ensemble (see \S~\ref{subsubsec:DistributionPk})
$
\mathscr{P}_{k}^\gc(n)\propto\EXP{-n\beta\varepsilon_k}
$
with $\varphi=1$.
This supports the fact that the condensed bosons in the ground state play the role of a reservoir for the excited bosons in this regime~\cite{Pol96} (cf. Appendix~\ref{app:GC}).

\begin{figure}[!ht]
 \centering
 \includegraphics[width=0.4\textwidth]{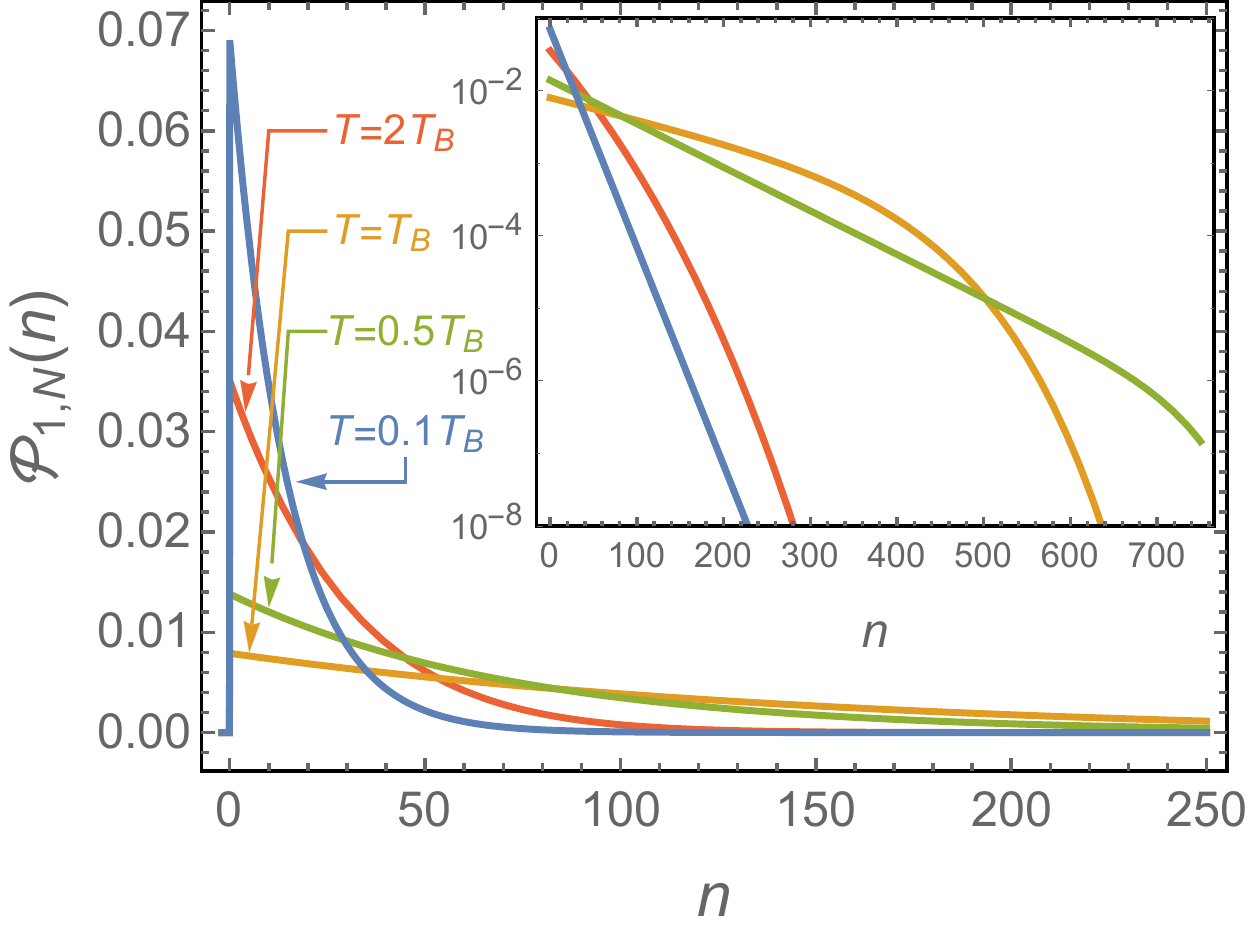}
 \caption{(Color line) 
 \textit{ Distribution $\mathscr{P}_{1,N}(n)$ of the number of bosons in the first excited state for $N=1000$. Temperature is $T/\TB=2$ (red), $1$ (orange), $0.5$ (green) and $0.1$ (blue), obtained from Eq.~\eqref{eq:DistributionNk}.}
 Inset~: 
 \textit{The plot in log-linear scale shows that the distribution is far from exponential when $T\gtrsim\TB$.
} }
 \label{fig:DitribN1}
 \end{figure}

For $ T \gtrsim \TB$, the distribution presents a decay faster than exponential (see Fig.~\ref{fig:DitribN1}). 

\subsubsection{Correlations}

We can also study the correlations between occupation numbers. For example, using \eqref{eq:UsefulRecurrence} we can easily get $\overline{n_kn_0}$.
Let us study the $T\to0$ limit of the correlator, when $\overline{n_0}\simeq N$ and $\overline{n_k}\simeq1/(\beta\varepsilon_k)$. 
In the continuum limit ($\beta\omega\ll1$) and for small enough $k$, we can expand the exponential $\EXP{\beta\varepsilon_k}\simeq1+\beta\varepsilon_k$. As a result we obtain 
$\mathrm{Cov}(n_0,n_k)=\overline{n_kn_0}-\overline{n_k}\times\overline{n_0}\simeq-\left( \overline{n_k} \right)^2$.
Denoting $N_\mathrm{ex}=\sum_{k>0}n_k$ the number of excited bosons, this result implies that 
$\mathrm{Cov}(n_0,N_\mathrm{ex})
\simeq-\sum_{k>0}\left( \overline{n_k} \right)^2
\simeq-(T/\omega)^2\sum_{k>0}1/k^2=-(1/6)(\pi T/\omega)^2\simeq-\mathrm{Var}(n_0)$, as it should since $\mathrm{Cov}(n_0,N_\mathrm{ex})=-\mathrm{Var}(n_0)=-\mathrm{Var}(N_\mathrm{ex})$ follows from the constraint that
$n_0+N_\mathrm{ex}=N$ is fixed.
Furthermore, we see that the anticorrelations
\begin{equation}
  \label{eq:CovNkN0}
  \frac{\overline{n_kn_0}-\overline{n_k}\times\overline{n_0}}{\sqrt{\mathrm{Var}(n_k)\mathrm{Var}(n_0)}}
  \simeq-\sqrt{\frac{\mathrm{Var}(n_k)}{\mathrm{Var}(n_0)}}
  \simeq - \frac{\sqrt{6}}{\pi\,k}
  \hspace{0.25cm}\mbox{for } 
  T\ll \TB
  \:,
\end{equation}
decay as higher excited states are considered.


\section{Conclusion}
     
We have obtained several general results for the occupation numbers in the \textit{canonical} ensemble for bosons and for fermions~: mean occupations,  fluctuations and correlation functions.
We have shown that the $p$-point correlation function for $N$ particles is expressed in terms of the $k$-body canonical partition functions, with $k=1,\ldots,N$, where these partition functions can be obtained by using a well-known recursion formula.
We have also obtained a representation of the $p$-point correlation function in terms of the ratio of two determinants, involving the mean occupations, which can therefore be viewed as the only fundamental quantities controlling any correlation function.
An open question would be to extend our determinantal representation to correlation functions involving arbitrary powers (in the bosonic case) and clarify the connection with the theory of symmetric functions in this case.

The two-point correlation function and the relation \eqref{eq:UsefulRecurrence} have recently found an application in Ref.~\cite{GraMajSchTex18}, where the variance of a specific observable for a gas of non-interacting fermions in a 1D harmonic trap was analysed in detail.
We have demonstrated the efficiency of our results by deriving some analytical expressions for the problem of Bose-Einstein condensation in a one-dimensional gas harmonically trapped. 
We have obtained significant deviations with the results given by the traditional grand canonical treatment where the constraint on the number of bosons is introduced \textit{a posteriori} (cf. Appendix~\ref{app:GC}).
This detailed analysis has relied on the knowledge of the exact canonical partition function. A study of higher dimensions or other situations would be interesting.

We have demonstrated that, in the regime where quantum correlations dominate ($T_Q=\omega\ll T\ll T_*=N\omega$), the distribution of the individual ground state occupancy has the form of a truncated Gumbel law.
Moreover, in the regime $T\ll\TB$, we get the Gumbel distribution.
Interestingly, this is not the first time that a connection is established between thermodynamical properties of a Bose gas and extreme value statistics~:
in Ref.~\cite{ComLebMaj07}, the spectral density of a Bose gas (not necessarily harmonically confined) was shown to be related to the different extreme value distributions for identical and independently distributed random variables. 
Depending on the exponent controlling the single-particle density of states $\rho(\varepsilon)\propto\varepsilon^{\alpha-1}$, the different universality classes (Gumbel, Fr\'echet or Weibull) can be obtained.
For the 1D harmonically trapped Bose gased studied here, the connection between the ground state occupancy distribution and extreme value statistics still remains to be explained.
     

\section*{Acknowledgements}

We acknowledge stimulating discussions with Jean-No\"el Fuchs, Satya Majumdar, Dmitry Petrov, Guillaume Roux and Gr\'egory Schehr.
We thank Kurt Sch\"onhammer for having pointed to our attention Ref.~\cite{Sch17}.


\appendix

\section{Schur functions}
\label{app:Schur}

Consider the integer partition $\lambda=(\lambda_1,\lambda_2,\ldots,\lambda_n)$ with $\lambda_1\geq\lambda_2\geq\cdots\geq\lambda_n$. If an integer is repeated, we may use the notation $(2,1,1,1)\equiv(2,1^3)$, here for the partition of $5$.
We introduce the specific partition $\delta=(n-1,n-2,\ldots,1,0)$.
Addition of partitions is simply obtained by adding integers term by term
$\lambda+\delta=(\lambda_1+n-1,\lambda_2+n-2,\ldots,\lambda_{n-1}+1,\lambda_n)$.
We introduce the determinant
\begin{equation}
  a_\lambda(x_1,\ldots,x_n)
  =\left|
  \begin{array}{cccc}
    x_1^{\lambda_1} & x_1^{\lambda_2} & \cdots & x_1^{\lambda_n} \\
    \vdots          &  \vdots         & \ddots & \vdots \\
    x_n^{\lambda_1} & x_n^{\lambda_2} & \cdots & x_n^{\lambda_n} 
  \end{array}
  \right|
\end{equation}
The Vandermonde determinant is then $a_\delta(x_1,\ldots,x_n)$, up to a sign.
The Schur function is defined by \cite{Mac95}
\begin{equation}
  s_\lambda(x_1,\ldots,x_n)
  = \frac{a_{\lambda+\delta}(x_1,\ldots,x_n)}{a_\delta(x_1,\ldots,x_n)}
  \:.
\end{equation}
Two examples are~\cite{Mac95}~: 
$
s_{(n,0^{n-1})}(x_1,\ldots,x_n)=h_n(x_1,\ldots,x_n)
$
and
$
  s_{(1^n)}(x_1,\ldots,x_n)=e_n(x_1,\ldots,x_n)=x_1 x_2\cdots x_n
$.


\section{Grand canonical treatment of bosons in a 1D harmonic trap}
\label{app:GC}

\begin{figure}[!ht]
\centering
\includegraphics[width=0.4\textwidth]{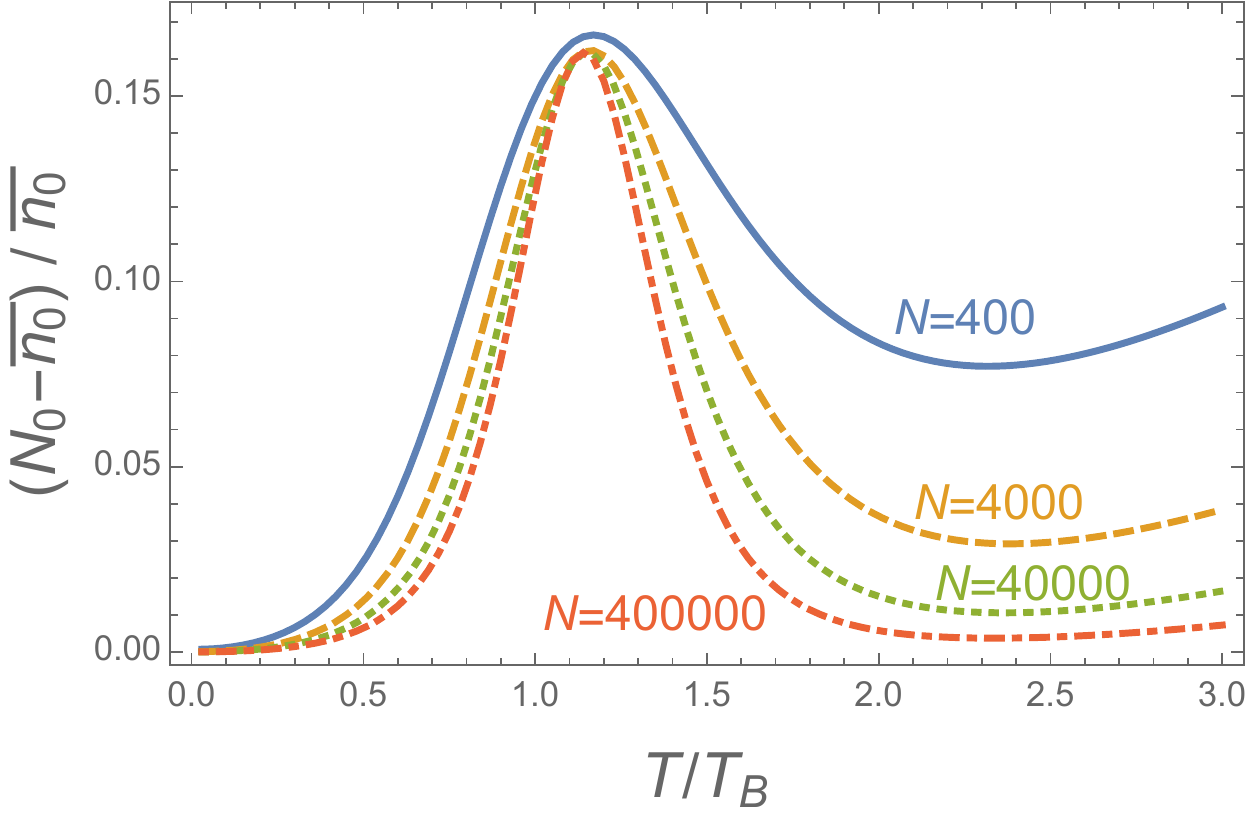}
\caption{\it
  Comparison between the form \eqref{eq:NKintegral}, obtained within the canonical treatment, and the result of the grand canonical treatment, solution of \eqref{eq:KetterleVanDruten}.}
\label{fig:N0ComparisonCanGrandcan}
\end{figure}

In this appendix, we recall the grand canonical treatment for bosons in a harmonic trap.
A first rough description can be found in \cite{Mul97,PetSmi02}, which corresponds to slightly adapt the usual treatment valid in $d>1$ \cite{PatBea11,TexRou17book}~:
while in $d>1$ the fugacity reaches $\varphi=1$ at the Bose-Einstein temperature, one needs to introduce a cutoff in 1D and set $\varphi=1-1/N$. 
This leads to the linear behavior~\cite{footnote3}. 
$\overline{n_0}/N\simeq1-T/\TB$ for $T<\TB$.
In pratice, the linear behavior is only reached for huge numbers of bosons because the fluctuation region is rather large in 1D \cite{PetGanShl04}.
A refined treatment was proposed in Ref.~\cite{KetDru96} (see also \cite{PetGanShl04})~:
assuming that the occupations are given by the usual Bose-Einstein factor \eqref{eq:BEandFD}, one splits the sum
 $N=\sum_\lambda\overline{n_\lambda}^\gc$   
between strongly occupied low energy levels and weakly occupied high energy levels. 
This leads to the equation for the condensate fraction \cite{KetDru96,PetGanShl04}~:
\begin{equation}
  \label{eq:KetterleVanDruten}
   N - \frac{T}{\omega}\,\ln(T/\omega) = N_0 - \frac{T}{\omega}\,
   \psi\left(1+\frac{T}{N_0\omega}\right)
\end{equation}
where $\psi(z)$ is the digamma function (we use a different notation for the (exact) canonical condensate fraction $\overline{n_0}$ and its counterpart $N_0$ in the (approximate) grand canonical approach).
From this equation, it is possible to recover the limiting behavior \eqref{eq:pseudoBEC}.
However a precise comparison with \eqref{eq:NKintegral} shows a relative difference of $\sim16\%$ when  $T\sim \TB$, irrespectively how large $N$ is, as shown in Fig.~\ref{fig:N0ComparisonCanGrandcan} (note that this discrepancy cannot be attributed to the continuous approximation leading to \eqref{eq:NKintegral}, cf. inset of Fig.~\ref{fig:1Dcondensate}).

\begin{thebibliography}{100}
\bibitem{footnote1}
Several simple thermodynamic properties (energy, heat capacity) are discussed in Ref.~\cite{TexRou17book} (see also Ref.~\cite{GraMajSchTex18}).
\bibitem{footnote2}
The method of Ref.~\cite{HolKalKir98} is suitable to analyse the moments of the total number of excited bosons and cannot be extended to study occupations of excited levels.
\end{thebibliography}
\end{document}